\documentclass[12pt]{iopart}

\usepackage{iopams}
\usepackage{graphicx}

\newcommand{\eqref}[1]{(\ref{#1})}

\newcommand{\bp}{{\bf p}}
\newcommand{\bq}{{\bf q}}
\newcommand{\bx}{{\bf x}}
\newcommand{\bX}{{\bf X}}
\newcommand{\bv}{{\bf v}}

\newcommand{\bu}{{\bf u}}
\newcommand{\bsig}{\hat{\boldsymbol \sigma}}

\newcommand{\sn}{\sigma_\nu}
\newcommand{\sm}{\sigma_\mu}
\newcommand{\sx}{\sigma_x}
\newcommand{\sy}{\sigma_y}

\graphicspath{{Figures/}}

\begin{document}

%%%%%%%%%%%%%%%%%%%%%%%%%%%%%%%%%%%%%%%%%%%%%%%%%%%%%%%%%%%%%%%%%%%%%%%%%%
\title[Violation of the Einstein relation in
Granular Fluids: the role of correlations]{Violation of the Einstein relation in
Granular Fluids: the role of correlations}
%\date{\today}

\author{A Puglisi$^1$, A Baldassarri$^1$ and A Vulpiani$^2$}
\address{$^1$ Dipartimento di Fisica, Universit\`a La Sapienza, p.le Aldo Moro 2, 00185 Roma, Italy}
\address{$^2$ Dipartimento di Fisica and INFN, Universit\`a La Sapienza, p.le Aldo Moro 2, 00185 Roma, Italy}

\ead{andrea.puglisi@roma1.infn.it}

\begin{abstract}
We study the linear response in different models of driven granular
gases.  In some situations, even if the the velocity statistics can be
strongly non-Gaussian, we do not observe appreciable violations of the
Einstein formula for diffusion versus mobility. The situation changes
when strong correlations between velocities and density are present: in
this case, although a form of fluctuation-dissipation relation holds,
the differential velocity response of a particle and its velocity
self-correlation are no more proportional. This happens at high
densities and strong inelasticities, but still in the fluid-like (and
ergodic) regime.
\end{abstract}

\maketitle

%%%%%%%%%%%%%%%%%%%%%%%%
%%%%%%%%%%%%%%%%%%%%%%%%
%%%%%%%%%%%%%%%%%%%%%%%%
\section{Introduction}

The transport properties of flowing dilute granular materials
constitute an open problem in non-equilibrium statistical
mechanics~\cite{C90,PL03}. The existing kinetic theories aiming to deduce
transport properties from the microscopic dynamics for the ``usual''
gases have a hard life, here, because of the presence of inelastic
interactions among grains, that prevent the assumption of an
equilibrium measure in the unperturbed state. Several approaches to
this problem have been proposed in previous studies.

In a large series of works, it has been considered a setup where
energy is injected only through the boundaries of the system. In this
treatment, the bulk is considered as a ``freely evolving'' inelastic
gas, which would cool down if not driven by energy currents
transported by the gas itself. It is possible to write down balance
equations for local fields such as density, velocity and kinetic
temperature and, through a very delicate assumption of separation
between microscopic and mesoscopic scales~\cite{G99}, the so-called
granular hydrodynamics can be obtained~\cite{BDKS98}. Following these
lines, fluctuation-response relations have been obtained with respect
to the homogeneous cooling state, i.e. relaxation laws for small
perturbations of a state whose fate is thermal death~\cite{DG01,DB02}.

A different approach consists in considering an alternative
experimental setup where the starting state is much more similar to a
thermal state, such as the thermodynamic equilibrium of a gas.  Such a
state can be prepared by coupling the energy source to all grains of
the system, for example in granular materials fluidized by some air
flow~\cite{LBLG00}, or otherwise in granular beds put on a vibrating
plate~\cite{OU98,PEU02}. In such cases, the unperturbed fluid state is
stationary and one can study how the system relaxes to it when a small
perturbation is applied. Models for these granular stationary states
have been proposed~\cite{WM96,PLMPV98,NETP99}, showing the main
differences with respect to the thermal state of a molecular gas: lack
of equipartition, departure from Gaussian statistics of velocities,
tendency to enhance spatial grain-grain correlations, clustering. It
must be remarked that the obtained stationary state is intrinsically
out of equilibrium: a net current of energy flows from the external
source, through inelastic collisions, into
heat~\cite{PVBTW05,VPBTW06}.  A Boltzmann equation for such class of
granular fluids has been proposed~\cite{NE98}, as well as a kinetic
theory for transport coefficients~\cite{BSSS99,GM02}: this theories
have their validity in a suitable range of the parameters. Recently numerical studies have been performed showing that, in
homogeneous situations, the Fluctuation-Response relation (FR) is
valid in its near-equilibrium formulation, replacing the bath
temperature with the internal granular
temperature~\cite{PBL02,BBDLMP05}. This has interesting consequences
in the case of mixtures, where different components have different
temperatures ~\cite{BLP04}: for instance,  a linear
response experiment on a massive tracer, performed to obtain a
temperature measurement (a granular thermometer), yields the
temperature of the tracer and not that of the surrounding gas. The
verification of FR has been explained by means of a hydrodynamic
approach by Garzo~\cite{G04}, who connected it to the very small
departures from the Maxwell-Boltzmann statistics. In the following, we
 discuss when this particular kind of Fluctuation Response
relation ceases to be valid in a driven granular system: it will appear
that the most relevant ingredient is not the deviation from the
Maxwell-Boltzmann statistics, but the degree of correlations among
different degrees of freedom (d.o.f.), which increases as the total
excluded volume decreases.

The aim of this paper is to put this problem in the more general
context of linear response theory for statistically stationary states,
whose formulation has been given in~\cite{DH75,FIV90} and which can
be described in very general terms. Consider a dynamical system $ {\bf
X}(0) \to {\bf X}(t)=U^t {\bf X}(0)$ whose time evolution can also be
not completely deterministic (\textit{e.g.}  stochastic differential
equations), with states ${\bf X}$ belonging to a $N$-dimensional
vector space.  We assume a) the existence of an invariant probability
distribution $\rho({\bf X})$, for which an ``absolute continuity''
 condition is required, and b) the mixing character of the
system (from which its ergodicity follows). These assumptions imply
also that the system is time translation invariant (TTI). Now we
introduce the two main ingredients of the theory: the response of the
system to a small perturbation, and the time correlation of the
unperturbed system that describes the relaxation of its spontaneous
fluctuations.  In the following we will indicate with $\langle \cdot
\rangle$ an average in the unperturbed system, i.e. weighting states
with the invariant measure, and with $\overline{(\cdot)}$ the time
dependent average in the dynamical ensemble generated by the external
perturbation.

In the unperturbed system, the relaxation of spontaneous fluctuations
is described by the time dependent cross correlations of two
generic observables $A(\bX)$ and $B(\bX)$
\begin{equation}
\label{eq:correlation_def}
C_{AB}(t)=\langle A(\bX(t)) B(\bX(0)) \rangle.
\end{equation}
The average effect at time $t$ on a variable $X_i$ of a small
external perturbation $f(s)$, for instance on the variable $X_j$, applied at time $s$ can be written, in the
linear response regime, as:
\begin{equation}
\label{eq:response_def2}
\overline{\delta X_{i}(t) }=\int_0^t R_{i,j}(t-s) f(s)ds \,,
\end{equation} 
which defines the response function $R_{i,j}(t)$. The basic idea of the
FR is to link the response functions $\{R\}$ to suitable correlations
$\{C\}$ of the unperturbed system.

For example, let us consider  a colloidal particle in a
fluid with friction constant $\gamma$ and temperature $T$: one has a
system with two d.o.f., position and velocity of the
particle, $(X,V)$, whose evolution is given by the Langevin equation:
\begin{eqnarray} \label{eq:langevin}
\frac{dX}{dt}&=&V\\
\frac{dV}{dt}&=&-\gamma V +\sqrt{\frac{2\gamma T}{M}}\eta,
\end{eqnarray}
where $\eta$ is a white noise, i.e. a Gaussian stochastic process with
$\langle \eta(t)\rangle=0$ and $\langle
\eta(t)\eta(t')\rangle=\delta(t-t')$ (for the sake of simplicity we
will assume the Boltzmann constant $k_B=1$). The time self-correlation
of the particle velocity, $C_{VV}(t)=\langle V(t)V(0) \rangle=\langle
V^2 \rangle e^{-\gamma t}$, when integrated from $0$ to $\infty$,
determines the self-diffusion coefficient: $D=\int_0^\infty \langle
V(t)V(0) \rangle dt$. It describes the asymptotic growth of the mean
square displacement of the particle: $\langle (X(t)-X(0))^2/t \rangle
\to 2D$. On the other side, when the momentum of the particle is
perturbed with a force $f(t)=F\Theta(t)$, where $\Theta(t)$ is the
Heaviside step function, the response of the velocity itself, at very
large times, reads:
\begin{equation}
\overline{\delta V(\infty)}=\frac{F}{M}\int_0^\infty R(t) dt=\mu F,
\end{equation}
which defines the mobility $\mu$. An easy computation gives
\begin{equation}\label{eq:mu_d}
\mu=\beta D,
\end{equation}
where $\beta=1/{T}$. This relation, well known as the Einstein
formula, obtained in his celebrated 1905 paper on Brownian
motion~\cite{E05,E06}, is a primordial example of the Fluctuation
Response relation: it relates, in fact, the response to a perturbation
with the relaxation of spontaneous fluctuations.

After the publication of the Einstein
relation, a large amount of work~\cite{O31,N28,G54,K57,K66} was
devoted to generalize it to the class of (classical as well as
quantal) Hamiltonian systems coupled to a thermostat at temperature
$T$. This means considering dynamical systems with variables
$(\bq,\bp)$ whose time evolution is generated by a Hamiltonian
$\mathcal{H}_0$. The external perturbation appears as a perturbation
of the Hamiltonian $\Delta \mathcal{H}(t)=f(t) B_f(\bq,\bp)$. In this
case, the Fluctuation Response relation can be written, among the
others, in the following form:
\begin{equation}
\frac{\delta A(t)}{\delta f(s)}=R_{A,f} (t) =\beta {\Bigl \langle A(t) \dot{B_f}(s) \Bigr \rangle } \;.
\end{equation}
The response of the observable $A$ at time $t$ with respect to a ``force''
$f$ applied to the system at time $s$ is related to the correlation,
measured in the unperturbed system, between the observable itself at
time $t$ and the time derivative of the observable $B_f$ at time
$s$. The latter is the one conjugated to  $f$ through the
Hamiltonian.

The fact that the FR theory was developed in the context of
equilibrium statistical mechanics of Hamiltonian systems generated
some confusion and misleading ideas on its validity.  As a matter of
fact it is possible to show that a generalized FR relation holds under
the rather general hypotheses discussed above, i.e. basically the
mixing property and the existence of an absolute continuous invariant
measure $\rho({\bf X})$.  The main result (for details
see~\cite{FIV90}) is the following Fluctuation-Response
relation, valid when considering the perturbation at time $0$ of a coordinate $X_j$:
\begin{equation}
\label{eq:fdt}
R_{i,j}(t) = \frac{\overline{\delta X_i(t)}}{\delta X_j(0)}=- \Biggl \langle X_i(t) \left.
 \frac{\partial \ln \rho({\bf X})} {\partial X_j} \right|_{t=0}
\Biggr  \rangle \, .
\end{equation}
From this relation all previous cases can be obtained. The Brownian
motion of the colloidal particle, for example, has an invariant
measure where position and velocity are independent: the part
concerning $V$ is of course a Gaussian with $\langle V \rangle=0$ and
$\langle V^2 \rangle=\frac{T}{M}$. From formula~\eqref{eq:fdt},
therefore, follows that 
\begin{equation} \label{eq:einstein}
R_{V,V}=M\beta \langle V(t) V(0) \rangle
\end{equation}
and this immediately returns the Einstein relation~\eqref{eq:mu_d}. In the
rest of the paper, with a slight abuse of terminology, we will use the
form ``Einstein relation'' to denote the time dependent form~\eqref{eq:einstein}.

In the case of thermostatted Hamiltonian systems, on the other side,
one has that $\rho(\bq,\bp) \propto \exp(-\beta
\mathcal{H}(\bq,\bp))$. In such case equation~\eqref{eq:fdt} gives for
example
\begin{equation}
R_{p_i,p_i} (t) =\beta {\Bigl \langle p_i(t) \left. \frac{\partial
\mathcal{H}}{\partial p_i}\right|_{t=0} \Bigr \rangle }=\beta {\Bigl \langle p_i(t) \frac{d}{dt}q_i(0) \Bigr \rangle }=-\beta \frac{d}{dt}{\Bigl \langle p_i(t) q_i(0) \Bigr \rangle }\; .
\end{equation}

In non Hamiltonian (and in general non-Gaussian) systems, the shape of
$\rho({\bf x})$ is not known, therefore~\eqref{eq:fdt} does not give a
straightforward information.  However from it one can see that a FR
relation still exists, stating the equivalence of the response to a
suitable correlation function computed in the non perturbed
systems. This mean that, from an ansatz on the invariant measure
$\rho$ one can directly deduce  the response matrix.

Following these lines, we analyze the response to small perturbations
of a thermostatted granular gas, trying to connect the response
properties of the stationary state with its many ``anomalies'' with
respect to an equilibrium state. In particular we show (section 2)
that, in homogeneous situations, even when the invariant measure is
far from the Gaussian, the Einstein relation holds with good
accuracy. On the contrary (section 3), when the granular effects
(excluded volume and inelasticity) are strong enough to develop
correlations between local density and velocities, the invariant
measure of the system becomes highly non-trivial, and the Einstein
relation is no more observed. We stress that the regimes considered
here are always ergodic: this is a relevant difference with respect to
previous studies on the violations of the Fluctuation-Response
relation, which considered glassy systems in the non-ergodic (aging)
phase~\cite{bckm98}.

%%%%%%%%%%%%%%%%%%%%%%%%
\section{Non Gaussian cases for which the Einstein relation holds}

\subsection{The models}

We start to discuss the linear response of a dilute granular gas with
$N$ grains of mass $m=1$. Three different models, all in dimension
$d=2$, are considered here:

\begin{enumerate}

\item the homogeneously driven gas of inelastic hard disks in the
dilute limit, evolving through stochastic Molecular Dynamics rules.

\item the homogeneously driven gas of inelastic hard disks in the
Molecular Chaos approximation, i.e. where its dynamics is determined
by Direct Simulation Monte Carlo (DSMC) algorithm

\item the inelastic Maxwell model driven by a ``Gaussian thermostat''

\end{enumerate}

In all the above models one has
\begin{equation} \label{eq:indep}
\rho(\{ \mathbf v_i, x_i\})=n^N\prod_{i=1}^N \prod_{\alpha=1}^d p_v(v_i^{(\alpha)})
\end{equation}
with $n$ the spatial density $n=N/V$ and $p_v(v)$ the one-particle
velocity component probability density function, $v_i^{(\alpha)}$ the
$\alpha$-th component of the velocity of the $i$-th particle and $d$
the system dimensionality. In particular, in models 2 and 3 this is
true by assumption, while for model 1 it is well verified in
simulations, as a consequence of being diluted. In view of the fact that
all discussed models are isotropic, in the following we will denote
with $v$ an arbitrary component of the velocity vector: the results do
not change if $v$ is the $x$ or $y$ component.

The three models are known to display non-Gaussian $p_v(v)$. 
From the above discussion, it is expected that
an instantaneous perturbation $\delta v(0)$ at time $t=0$ on a
particle of the gas, will result in an average response of the form
\begin{equation} \label{eq:fdt_1p}
R(t)=\frac{\overline{\delta v(t)}}{\delta v(0)}=-\left\langle \left . v(t)\frac{\partial \ln p_v(v)}{\partial v}\right |_0 \right\rangle \neq C_1(t),
\end{equation}
having defined $C_1(t)=\langle v(t) v(0)\rangle/\langle v^2\rangle$

Some previous studies already showed that for the inelastic hard disks
model, in the dilute limit, it is very difficult to observe the
discrepancy between $R(t)$ and $C_1(t)$, i.e. to see any ``violation''
of the Einstein relation for mobility and diffusion. In such studies,
however, the deviation from a Gaussian $p_v(v)$ was always small,
i.e. consistent with a Sonine polynomial fit with a parameter $a_2
=\frac{\langle v^4\rangle}{3 \langle v^2\rangle^2}-1 \ll
1$~\cite{NE98}. Actually, in a driven dilute system it is very rare to
observe large departures from the Gaussian behavior. On the other
side, studying models such as the thermostatted Maxwell model, or
tuning the parameters of the DSMC algorithm for inelastic hard disks
beyond the dilute limit, one can induce rather large deviations from
the Gaussian, while condition~\eqref{eq:indep} still holds. Even if
this may be far from being realistic, it is useful to assess the
relevance of the non-Gaussian velocity pdf on the linear response of
the gas.

For the three models, the unperturbed dynamics is determined by a
non-interacting streaming (where each particle is coupled to the
thermostat only) plus inelastic collisions. For model (i), which is
the closest to experiments of driven granular gases, the streaming
part is described by the equations of motions of $N$ hard disks of diameter $\sigma=1$
moving in a square of area $V=L \times L$ with periodic boundary
conditions and coupled to a thermal bath with viscosity $\gamma$ and temperature $T_b$:
\begin{eqnarray} \label{eq:hs}
\frac{d \bx_i(t)}{dt}&=&\bv_i(t)\\
\frac{d \bv_i(t)}{dt}&=&-\gamma \bv_i(t) + \sqrt{2\gamma T_b}\boldsymbol{\eta}_i(t), \label{eq:hs2}
\end{eqnarray}
with $\boldsymbol{\eta}$ a Gaussian white noise, i.e. $\langle
\eta_i^{(\alpha)}(t) \rangle=0$ and $\langle
\eta_i^{(\alpha)}(t)\eta_j^{(\alpha')}(t')\rangle=\delta_{\alpha,\alpha'}\delta_{ij}\delta(t-t')$,
where $\alpha$ and $\alpha'$ indicate the Cartesian components. When
two grains $i$ and $j$ touch, an instantaneous inelastic collision
takes place, with a change of velocities given by
\begin{equation} \label{eq:colrule}
\bv_i'=\bv_i-\frac{1+r}{2}[(\bv_i-\bv_j)\cdot \bsigma]\bsigma=\bv_i+\Delta \bv_{i,col}(\bv_i,\bv_j,\bsigma)
\end{equation}
where $r \in [0,1]$ is the restitution coefficient (the elastic case
corresponds to $r=1$) and $\bsigma$ is the unit vector joining the
centers of the two colliding particles. In the dilute limit, numerical
simulations show that colliding particles are not correlated,
i.e. Molecular Chaos holds. The system is known to display very
different regimes when $\tau_b/\tau_c$ changes, where
$\tau_b=1/\gamma$ and $\tau_c=1/\omega_c$ is the average mean free
time between collisions of a single particle. If $\tau_b/\tau_c \gg
1$, the effect of collisions is very small and, even if inelastic, the
gas behaves as at equilibrium at temperature $T_b$. In the opposite
case, $\tau_b/\tau_c \ll 1$, collisions are dominant and the gas
reaches a fluctuating stationary state with ``granular temperature''
$T_g \equiv \frac{\langle |v|^2 \rangle}{2} < T_b$ ($T_g$ is smaller,
the smaller $r$ is). In this non-equilibrium regime, the velocity pdf
is non-Gaussian with slow tails at very large $|v|$. Note that, here,
$\tau_c$ is not an external parameter, but is self-determined by the
system: increasing $n \sigma$, i.e. reducing the mean free path,
results in a smaller granular temperature and the same happens when
increasing $\tau_b$, therefore in both cases a direct increase of
$\tau_b/\tau_c$ is not obvious. A direct experience teaches that, when
diluteness (volume fraction $\phi= n \pi \sigma^2 /4 \ll 1$) is also
required, then it is very difficult to obtain $\tau_c \ll \tau_b$. For
instance, when $\phi \sim 0.1$, we usually observe $\tau_c \sim 0.1
\tau_b$ or larger.

Model (ii) consists of the same physical system but now the Molecular
Chaos assumption is artificially enforced. This is achieved by
disregarding the spatial coordinates of the particles, and selecting
with a stochastic rule the pairs of particles involved in each
collision. Time is discretized in small steps of length $\Delta t$
(smaller than $\tau_b$ and $\tau_c$). At each step the discretized
version of~\eqref{eq:hs2} is used to evolve the velocities of all the
particles. Then, a number of collisions $N\omega_c \Delta t/2$ is
performed, where $\omega_c=2\sigma n \sqrt{\pi T_g}$ is the
theoretical one-particle collision frequency for a dilute gas. Pairs
$i,j$ to collide are chosen with a probability proportional to the
quantity $-(\bv_i-\bv_j) \cdot \bsigma \Theta (-(\bv_i-\bv_j) \cdot
\bsigma)$, with $\bsigma=(\cos \theta, \sin \theta)$ and $\theta$
chosen randomly with uniform probability in $[0,2\pi)$. This process
mimics the relative velocity dependent collision frequency in dilute
gas of particles with hard core interactions. Since, typically, a particle $i$,
after having collided with a particle $j$, will do a second collision
with the same particle $j$ after a number of collisions of order $N$,
any memory of the first collision will be lost and the new collision
can be considered uncorrelated to the previous one. The phenomenology
observed in this model is analogous to that of model (i) in its dilute
limit. On the other side, here one can arbitrarily tune the ratio
$\tau_b/\tau_c$, increasing $n \sigma$ and ignoring the inconsistency
between the high density and the enforced Molecular Chaos.

Finally, model (iii), the inelastic Maxwell model, is a further
simplification of model (ii): couples of particles collide with a
constant probability, i.e. independently of the relative velocity. The
collision frequency is assumed to be $\omega_c=1/N$. Moreover, the
streaming part of the dynamics is performed with $T_b=0$ and
$\gamma=-\lambda$ with $\lambda=(1-r^2)/4$, i.e. a negative friction
and no random forces. This is the so-called Gaussian thermostat, which
guarantees a constant kinetic energy, since in the non-driven case
(``free cooling'') the total energy of the system would decay as $\sim
\exp(-\lambda t)$. It has been shown that such ``thermostat'' is
equivalent to consider the free cooling system and continuously
rescaling all particles' velocity components by a factor
$\sqrt{T_g}$. The analysis of this idealized granular model is
instructive for the following reasons. First, it has been
shown~\cite{BMP02} that such model has a stationary probability
density function for the velocity $bv$ with power law tails. In particular,
it displays $p_v(v)$ with high energy tails of the form $v^{-b}$ with
$b= 4$ in $d=1$ and $b>4$ in $d=2$ (a good estimate for not too high
inelasticity is $b \simeq 4/(1-r)$). Second, the 
simplification of the dynamics allows  a direct analytical
computation of time correlations and responses.

\subsection{The numerical experiment}

The protocol used in our numerical experiment, for the three models,
is the following:

\begin{itemize}

\item First, the gas is prepared in a ``thermal'' state,
with random velocity components extracted from a Gaussian with zero
average and variance $T_g(0)$. Positions of the particles, relevant
only in model (i), are chosen uniformly random in the box, avoiding
overlapping configurations. 

\item Second, the system is let evolve until a
statistically stationary state is reached, which is set as time $0$:
we verify that the total kinetic energy fluctuate around an average
steady value and that this value does not depend upon initial
conditions. In the case of model (iii) the energy is stationary by
definition, therefore we ensure that a stationary velocity pdf is
observed. 

\item Third, a copy of the system is obtained, identical to the
original but for one particle, whose $x$ (for instance) velocity component
is perturbed of an amount $\delta v(0)$. 

\item Finally both systems are let
evolve with the unperturbed dynamics. In models (i) and (ii), which
involve random thermostats, the same noise is used. After a time
$t_{max}$ large enough to have lost memory of the configuration at
time $0$, a new copy is done with perturbing a new random particle and
repeating the response measurement. 

\end{itemize}

This procedure is performed many
times, in order to reduce the statistical errors for both the response
and all required self-correlations in the unperturbed copy. In the
following, averages indicated as $\langle \cdot \rangle$ and
$\overline{(\cdot)}$ will have the meaning of averages over many
realizations of this procedure.

\begin{figure} 
\includegraphics[width=7.5cm,clip=true]{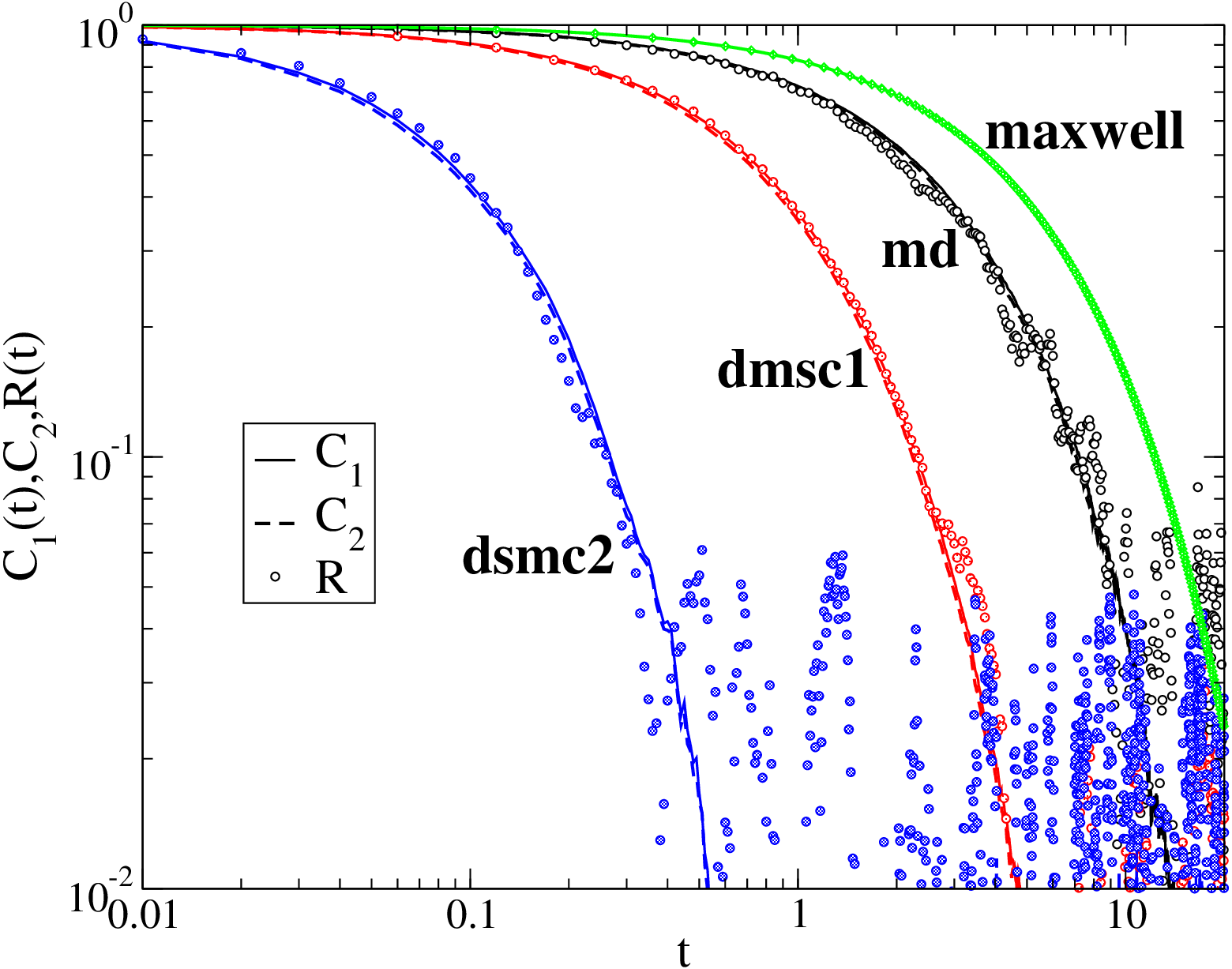}
\includegraphics[width=8cm,clip=true]{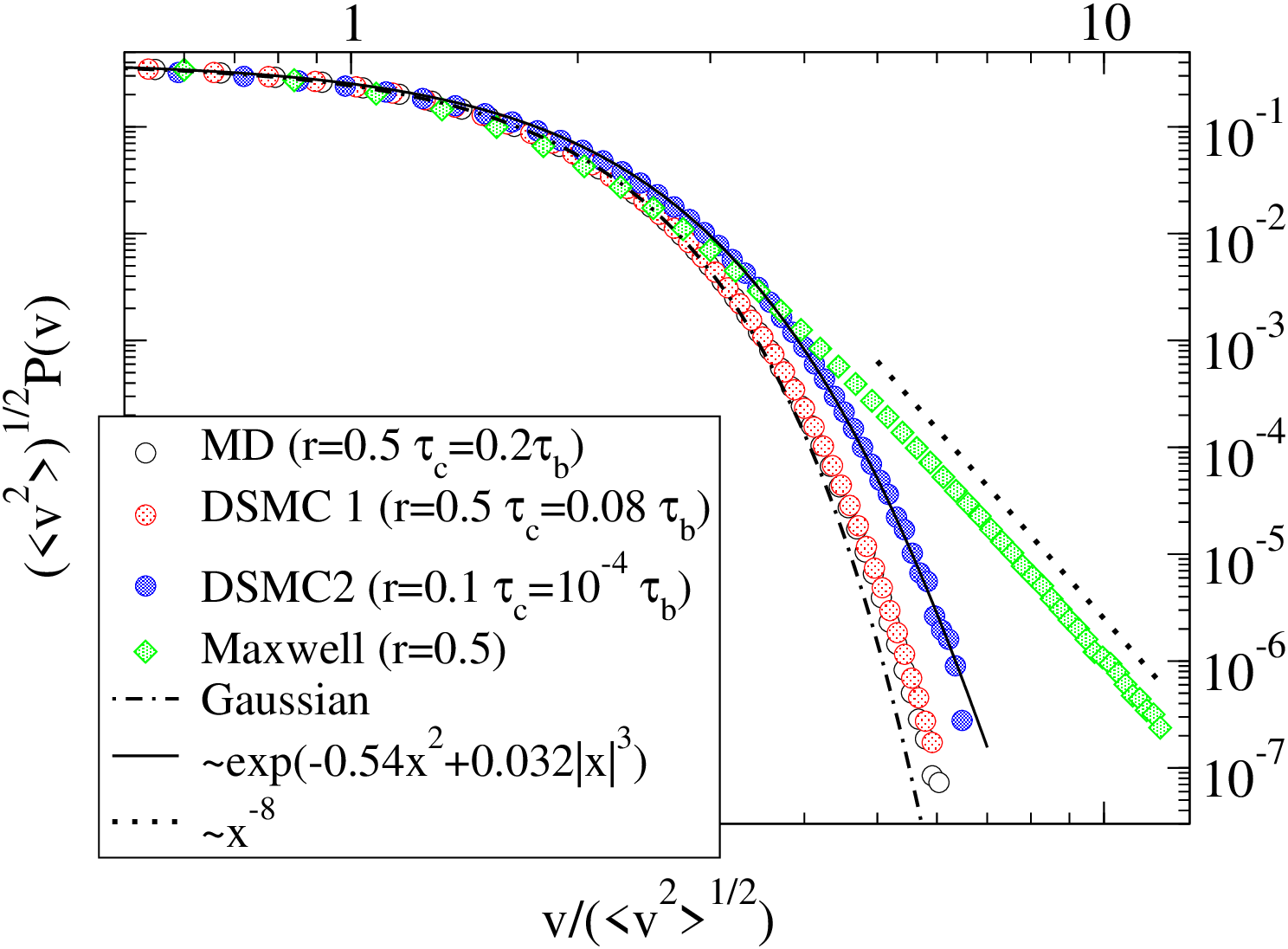}
\caption{Left: correlation functions $C_k(t)$ and response functions
$R(t)$ vs. time for different models described in the text. Right:
pdfs of the $x$-component of the velocity (here denoted as $v$), for
the same models. In all simulations $N=1000$. In the $MD$ simulation
the box is of size $100 \times 100$, $\tau_b=1/\gamma=10$ an
$T_b=1$. All other parameters are reported in the figure.
\label{fig:nocorr}}
\end{figure}

In Figure~\ref{fig:nocorr} we show the results of these experiments for
the three different models. In the right frame the velocity pdfs are
shown for different choices of the parameters in those models. In
Molecular Dynamics simulations of inelastic hard disks, even if quite
inelastic, but still dilute ($\phi=n \pi \sigma^2 /4 < 0.1$), the
velocity pdf is not very far from a Gaussian, as in a DSMC with similar
choices of the parameters ($\tau_c \sim 0.1 \tau_b$). Increasing $n$
in the DSMC leads to stationary regimes very far from thermal
equilibrium, with $T_g \ll T_b$ and larger tails of the velocity pdf
$p_v(v)$, with $a_2 \sim 0.1$. In view of relation~\eqref{eq:fdt_1p},
we have tried a three parameter fitting of the kind 
\begin{equation} \label{eq:fit}
p_v(v)=c_0 \exp(-c_1 v^2+c_2 |v^3|+c_3 v^4)
\end{equation}
where $c_0$ is not independent
because of normalization. In most of observed cases $|c_3 v_{max}^4|
\ll |c_2 v_{max}^3|$, with $v_{max}$ the largest value of $v$ in the
histogram. Therefore for our aims, in practice, we can drop the (negative) quartic term,
retaining only $c_1$ and $c_2$. The obtained fit appears to be very
good, see Figure~\ref{fig:nocorr}. Using this formula in
equation~\eqref{eq:fdt_1p}, we get
\begin{eqnarray} \label{eq:respsingle}
R(t)=-2c_1\langle v(t)v(0) \rangle+3c_2\langle v(t)|v(0)|v(0)\rangle\\=-2 c_1 \langle v^2 \rangle C_1(t)+ 3 c_2 \langle |v|^3 \rangle C_2(t),
\end{eqnarray}
which also defines $C_2(t)$. This is an example of the discussion
given in the introduction: an ansatz on the invariant measure leads to
a link between the response function and some correlation
functions. Note that here the ansatz is composed of two assumptions:
positions and velocities of the grains are independent (this is exactly true for models (ii) and
(iii)), equation~\eqref{eq:indep}, and a specific non-Gaussian shape
of $p_v(v)$. Finally, for the Inelastic Maxwell Model the tail of the
velocity pdf shows a power law decay with an exponent in agreement
with its quasi-elastic limit $4/(1-r)=8$ when $r=0.5$.

In the left frame of Figure~\ref{fig:nocorr}, we have superimposed
the response $\overline{\delta v(t)}/\delta v(0)$ to
the time self-correlations of different orders $C_1(t)$ and $C_2(t)$
measured in the unperturbed system. From the above results, we learn that in
all the considered models:
\begin{itemize}
\item 
different correlations are almost identical (we do not show $C_3(t)=\langle v(t)v^3(0)
\rangle/\langle v^4 \rangle$, but the result is very close)
\item 
a very good agreement between $R(t)$ and $C_1(t)$ is observed, equivalent
to a verification, within the limits of numerical precision, of the
Einstein relation.
\end{itemize}

The observation that the self-correlations $C_k(t)$, at least for
$k=1,2,3$, are almost identical is very robust. With a precise
statistics one can appreciate small differences at large times,
proving that it is not an exact equivalence. Anyway, the measurement
of response function is usually very noisy, and it is not easy to have
a good signal/noise ratio at such late times. Therefore, for the
practical purpose of the linear combination involved in the response,
these small differences are negligible and the Einstein relation is
practically satisfied.

It is interesting to note that a rather similar
situation is encountered when studying a different system, i.e. a gas
of non-interacting particles whose velocities obey a Langevin equation
with a non-quadratic potential:
\begin{equation}
\frac{d v(t)}{dt}=-\gamma\frac{d U(v)}{dv}+\sqrt{2 \gamma}\eta(t),
\end{equation}
with $U(v)=c_1 v^2-c_2 v|v|^2+c_3 v^4$ (with positive $c_1$, $c_2$ and
$c_3$). A numerical inspection, not shown here, clearly indicates that
$C_1(t)$, $C_2(t)$ and $C_3(t)$ are almost indistinguishable.

A simple condition can be given for the observed behavior. In fact, a
generic time correlation for $v(t)$ with a function $f[v(0)]$ can be
written as
\begin{eqnarray} \label{eq:angelo}
\langle v(t) f[v(0)] \rangle=\int dv_t \int dv_0 p_v(v_0) \mathcal{P}_t(v_t|v_0) v_t f(v_0)=\\\int dv_0 p_v(v_0) f(v_0) \langle v_t | v_0 \rangle,
\end{eqnarray}
where $\mathcal{P}_t(v_t|v_0)$ is the conditional probability of
observing $v(t)=v_t$ if $v(0)=v_0$ (time translation invariance is
assumed) and $\langle v_t | v_0 \rangle=\int dv_t
\mathcal{P}_t(v_t|v_0)v_t$ is the average of $v(t)$ conditioned to
$v(0)=v_0$. 

If, for some reason, $\langle v_t|v_0 \rangle=g(t)q(v_0)$, with $g$
and $q$ two given functions, then the dependence on $t$ results
independent of the choice of the function $f(v)$, i.e. on the order of
the correlation. This happens in model (iii), where in spite of the
non-Gaussian shape of the velocity pdf, the equivalence between $R(t)$
and $C_1(t)$ and any other correlation

\begin{equation}
C_f(t)=\frac{\langle v(t)f[v(0)]\rangle}{\langle v(0)f[v(0)]\rangle}=
R(t)=\exp\left(-\frac{r(r+1)}{4}t\right ),
\end{equation}
with any generic function
$f$ of the initial velocity value, is exact, see appendix B for the
case $d=2$ and~\cite{BBDLMP05} for $d=1$.

%%%%%%%%%%%%%%%%%%%%%%%%
\section{Non homogeneous granular fluids}

The factorization of the invariant phase space measure,
Eq.~\eqref{eq:indep}, is no more obvious in model (i) when density
increases. Correlations between different d.o.f., that is
positions and velocities of the same or of different particles, appear
also in homogeneously driven granular gases, as an effect of the
inelastic collisions that act similarly to an attractive
potential. Such a phenomenon has been discussed for this model of bath
in~\cite{PLMPV98,PLMV99,CDMP04} and for other homogeneous
thermostats in~\cite{WM96,NETP99,TPNE01}. In~\cite{PLMPV98,PLMV99}
it was also discussed the interplay between local density and local
granular temperature, which in some very dissipative cases present
strong fluctuations correlated to each other. These correlations
indicate a breakdown of the factorization of the invariant measure, in
particular at the level of velocity with respect to position of the
same particle. As a matter of fact, these effects result in a strong
violation of the Einstein relation, and in general of the equivalence
between $R(t)$ and $C_1(t)$.

Even in the presence of correlations, one can define and
compute the marginal probability density function of the component $x$ of the
velocity of one particle $i$, projecting the phase space measure $\rho(\{ \bv_i, \bx_i\})$:
\begin{equation}
f_i(v)=\int \prod_{k=1}^{N} d\bx_i \prod_{k=1,k\neq i}^N d\bv_k dv_i^y\rho(\{ \bv, \bx\}).
\end{equation}
However not necessarily this function has a role in the
response function. For example, perturbing the $x$ component of the
velocity of the $i$-th particle and measuring the response of the same
component, one obtains
\begin{equation} \label{eq:respmulti}
R(t) = - \Biggl \langle v_i^x(t) \left.  \frac{\partial \ln \rho(\{
 \bv, \bx\})} {\partial v_i^x} \right|_{t=0}\Biggr \rangle \ \neq - \Biggl
 \langle v_i^x(t) \left.  \frac{\partial \ln f_i(v_i^x)}
 {\partial v_i^x} \right|_{t=0} \Biggr \rangle \, .
\end{equation}

\begin{figure}
\includegraphics[width=8cm,clip=true]{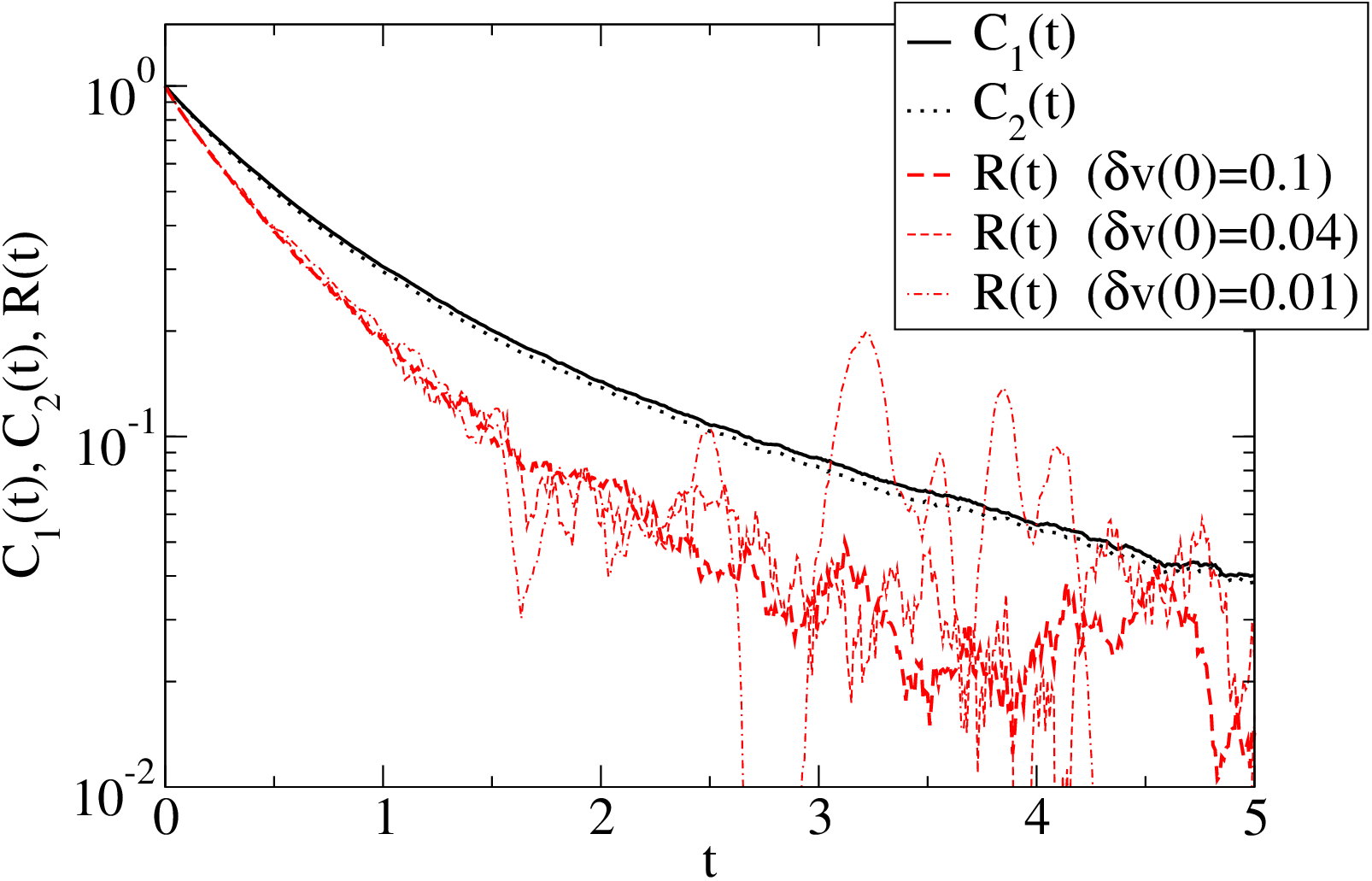}
\includegraphics[width=8cm,clip=true]{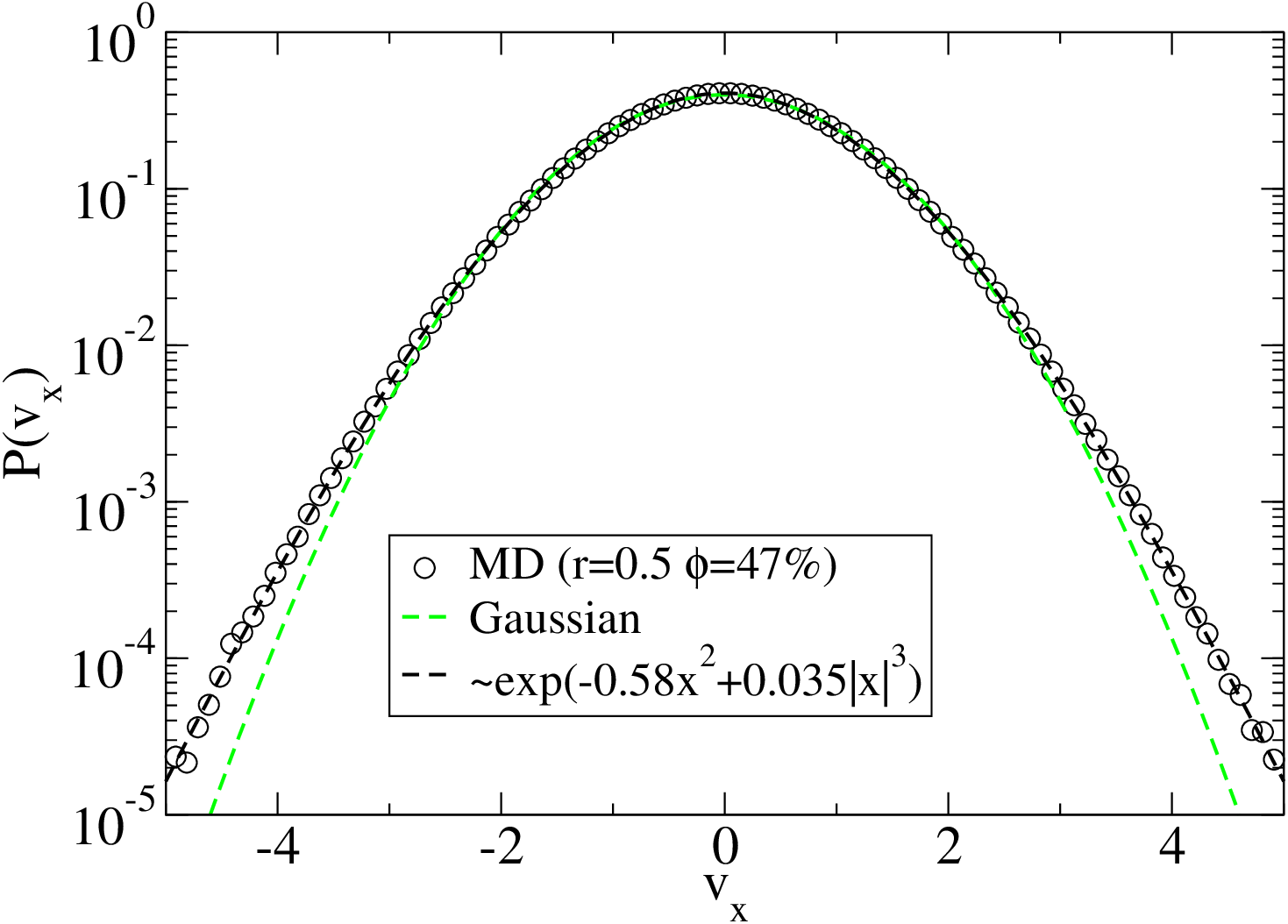}
\caption{Left: correlations functions $C_k(t)$ and response functions
$R(t)$ vs. time for a dense $MD$ simulation: the response function is
reported for different values of the perturbation $\delta
v(0)$. Right: pdf of the $x$ velocity component.  The system has
$N=1000$, box of size $41\times 41$, $\tau_b=1/\gamma=10$. In the
simulation the mean free time between collisions is measured to be $\tau_c=0.03 \tau_b$.
\label{fig:denso}}
\end{figure}

This is exactly what happens in model (i) when density is
increased. In Figure~\ref{fig:denso}, left frame, the correlation
functions $C_1(t)$ and $C_2(t)$ are shown, together with the response
function measured with different values of the perturbation $\delta
v(0)$. The very good agreement between different response functions
guarantees that the system is indeed linearly perturbed. At the same
time, the different correlations functions $C_k(t)$ are very close,
reproducing the phenomenology already observed in the previous dilute
cases, with the difference that the time dependence is not exponential
but slower, closer to a stretched exponential $\sim
\exp(-(t/\tau)^\alpha)$ with $\alpha<1$. Finally, looking at the
velocity pdf of the gas, the previously proposed exponential of a
cubic polynomial, Eq.~\eqref{eq:fit} with a negligible $c_3$
coefficient, is found to perfectly fit the numerical
results. Therefore, if the correlations among the different d.o.f. are
neglected, using equation~\eqref{eq:respsingle} and the
proportionality of the functions $C_k(t)$, a verification of the
Einstein formula $R(t) \simeq C_1(t)$ is still expected. The results
displayed in Figure~\ref{fig:denso}, left frame, demonstrate that this
is not the case: the hypothesis of weak correlations among different
d.o.f. must be dropped and the correct formula for the response is
Eq.~\eqref{eq:respmulti}. Unfortunately it is not very easy to use
such a relation.

\begin{center}
\begin{figure}
\includegraphics[width=12cm,clip=true]{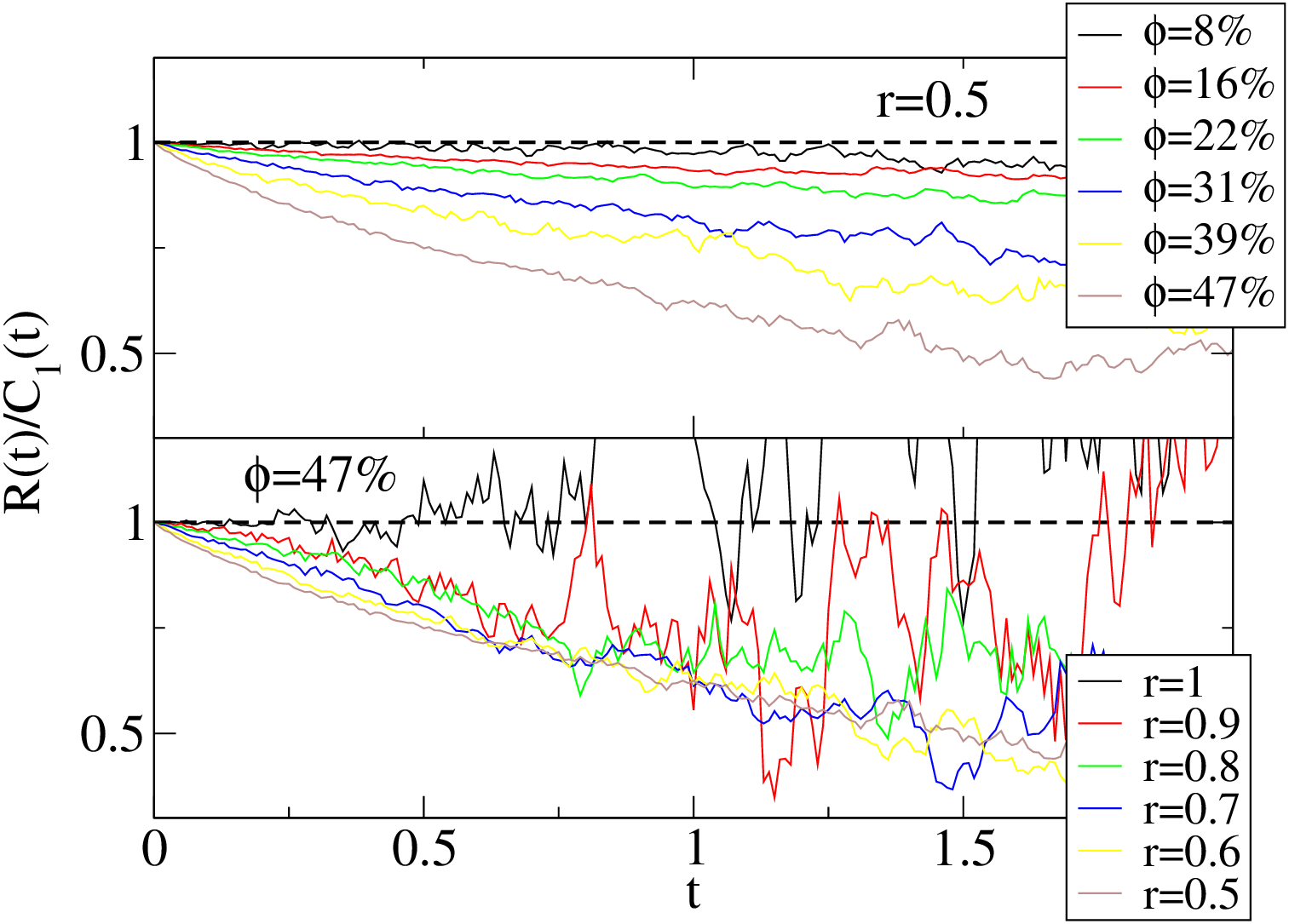}
\caption{Ratio between the response function $R(t)$ and the normalized
velocity self-correlation $C_1(t)$. The ratio is $1$ when the Einstein
relation is satisfied. All the results come from MD simulations with
$N=1000$ particles, $T_b=1$ and $\tau_b=1/\gamma=10$. Different values
of the covered fraction and of the restitution coefficient are used,
as shown in the figure.  \label{fig:molti}}
\end{figure}
\end{center}

\begin{center}
\begin{figure}
\includegraphics[width=12cm,clip=true]{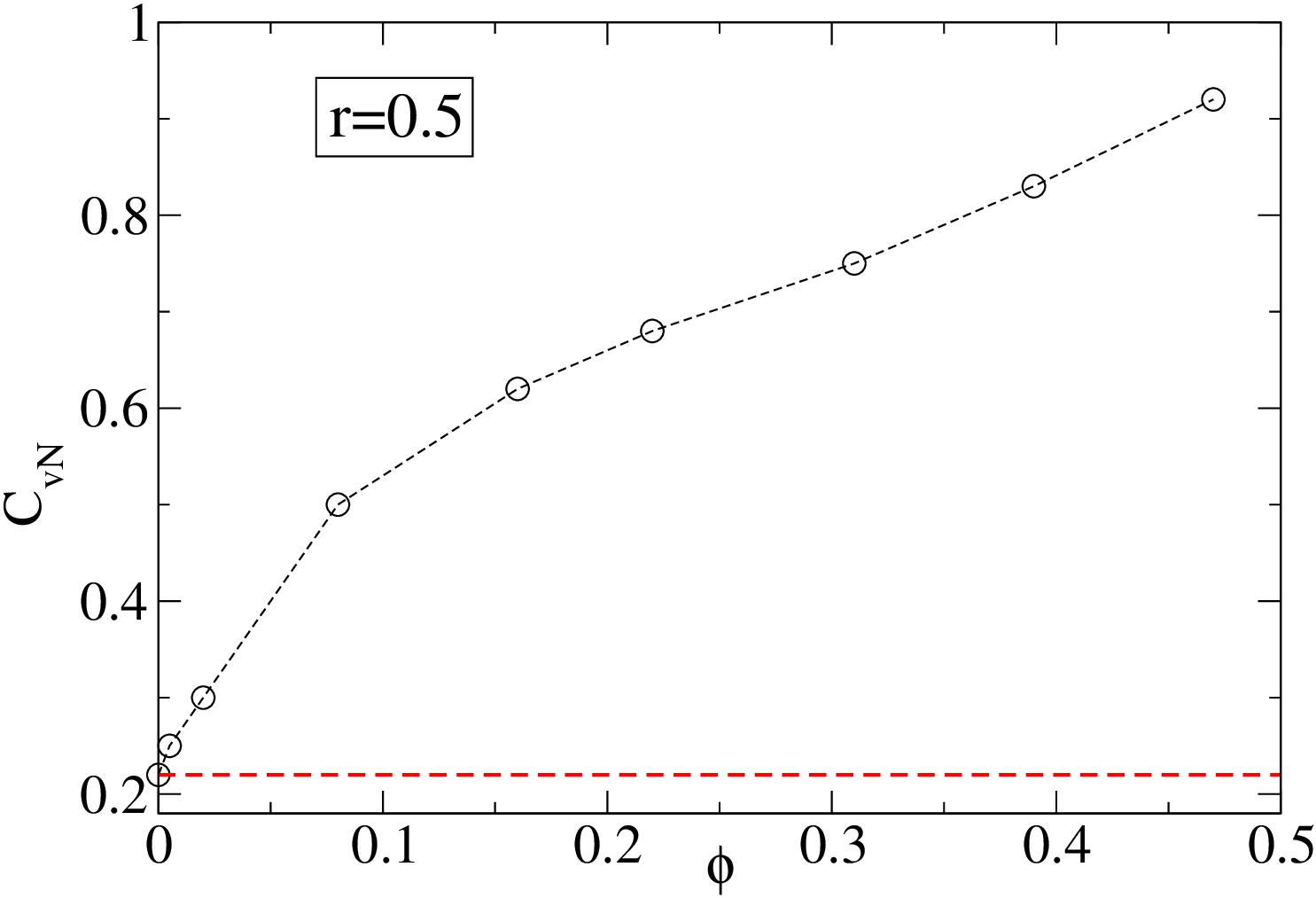}
\caption{Correlation between the square of the particle $x$ velocity
and local density $C_{vN}=(<v_i^2 N_i>-<v_i^2><N_i>)/(<v_i^2><N_i>)$
for different values of the volume fraction, in MD simulations with
$N=1000$, $T_b=1$ and $\tau_b=10$. In the dilute limit $\phi \to 0$,
$C_{vN} \to C_{vN}^*$, which is different from zero because of the
total finite number of particles. The red dashed line shows the
estimate of $C_{vN}^*$ obtained throwing $N$ random velocities,
extracted from a Gaussian distribution, into random boxes of the same
size used in the MD, and repeating the measure over many independent
realizations.
\label{fig:vdens}}
\end{figure}
\end{center}

The degree of violation of the Einstein formula increases with the
volume fraction $\phi$ and the inelasticity $1-r$, as shown in
Figure~\ref{fig:molti}, where we have reported the ratio $R(t)/C_1(t)$
as a function of time. This observation is consistent with the above
argument: correlations among different d.o.f. increase when the
probability of repeated contacts (the so-called ``ring collisions'')
is enhanced, and this happens when the excluded volume and/or the
post-collisional relative velocity are reduced. In the elastic case,
$r=1$, no violation is observed. A direct test of the existence of
non-trivial correlations in the system is given in
Figure~\ref{fig:vdens}. Each point in this figure represents the value
of $C_{vN}=(<v_i^2 N_i>-<v_i^2><N_i>)/(<v_i^2><N_i>)$ for a given
particle $i$, with $N_i$ the number of particles in a squared box
centered in $\bx_i$ and of size $L/15$, measured on a long trajectory
of the unperturbed system. We observe that $C_{vN}$ increases together
with the volume fraction $\phi$.
%Unfortunately the correlator $C_{vN}$
%depends almost only upon density, so that an analogous test with fixed
%$\phi$ and changing $r$ gives an almost constant value.

%%%%%%%%%%%%%%%%%%%%%%%%%%%%%%%%%%
\subsection{A Langevin model with two correlated variables}

In order to show in a clear way the role of correlations, we discuss now a simple model with only two variables:

\begin{eqnarray} \label{eq:lange}
\frac{dx(t)}{dt}&=&m_{11}x(t)+m_{12}v(t)+\sigma_{11} \eta_1(t)+\sigma_{12}\eta_2(t)\\
\frac{dv(t)}{dt}&=&m_{12}x(t)+m_{22}v(t)+\sigma_{21} \eta_1(t)+\sigma_{22}\eta_2(t)
\end{eqnarray}
If the matrices $\hat{m}$ and $\hat{\sigma}$ are diagonal, the two
variables are independent. Provided that the symmetric matrix
$\hat{m}$ has negative eigenvalues and $\det \hat{\sigma} \neq 0$, the
pdf of $(x,v)$ relaxes toward a bi-variate Gaussian function. Instead
of discussing the general form, we consider the case whose invariant
joint pdf is
\begin{equation} \label{eq:joint}
\rho(x,v) \propto \exp(-\frac{x^2}{2}-\frac{v^2}{2}+\frac{xv}{2}).
\end{equation}
Of course the marginal pdf of each single variable is a
Gaussian. Neglecting the correlation among $x$ and $v$, the response
of $v$ to a perturbation on itself, would again be expected to be equal to
$C_1(t)=\langle v(t) v(0) \rangle/\langle v^2 \rangle$. On the
contrary, the correct response is given using the full
formula~\eqref{eq:fdt} applied to the joint pdf~\eqref{eq:joint}. The result is

\begin{equation} 
R(t)=\langle v(t)v(0) \rangle-\frac{1}{2} \langle v(t) x(0) \rangle.
\end{equation}

\begin{center}
\begin{figure}
\includegraphics[width=12cm,clip=true]{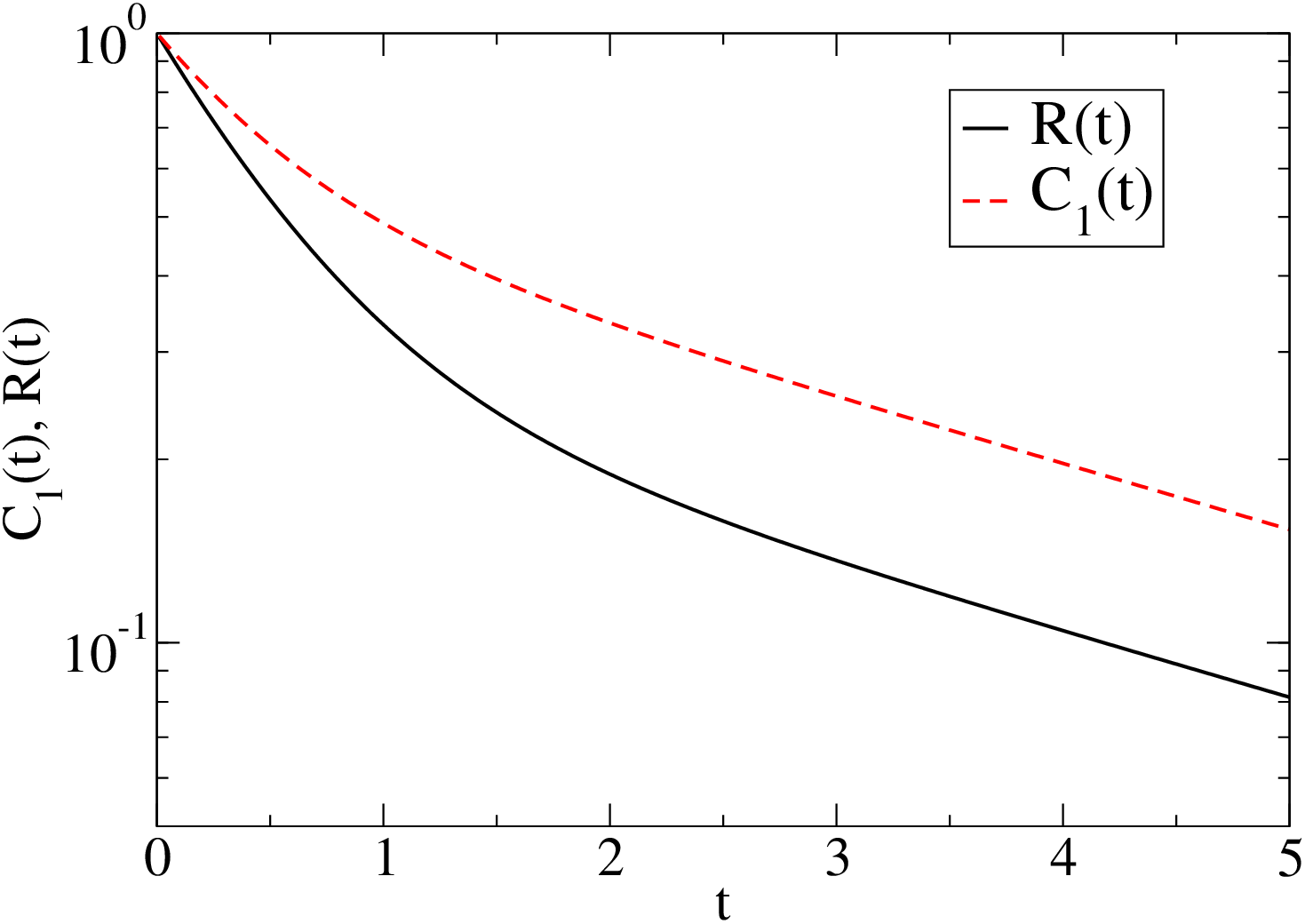}
\caption{Response $R(t)$ and velocity correlation $C_1(t)$ in the simple Langevin model with correlated variables discussed in Eq.~\eqref{eq:lange}, with parameters $m_{11}=-1.1$, $m_{12}=0.8$, $m_{22}=-1$. \label{fig:lange}}
\end{figure}
\end{center}

The difference between the Einstein formula and the correct response
is shown in Figure~\ref{fig:lange} for a choice of the matrix $\hat{m}$.

%%%%%%%%%%%%%%%%%%%%%%%%
\section{Conclusions}

In this paper we have reported the analysis of linear response in
different models of driven granular gases, which have the property of
rapidly reaching a statistically stationary state. The response
function is directly related to the global invariant measure in the
phase space, which is unknown for this kind of non-equilibrium
systems. When positions and velocities of the particles are not
correlated, the response of a perturbation on the velocity of a single
particle is expected to depend on the singlet velocity pdf, which can
be close or far from a Gaussian, depending on the model and on
physical parameters. Nevertheless, the existence of a unique time
scale that characterizes all possible correlation functions, makes the
exact form of the velocity pdf irrelevant for the response function:
the latter is, in practice, always indistinguishable from the
normalized velocity self-correlation $\langle v(t)v(0) \rangle/\langle
v^2 \rangle$. This is equivalent to say that the Einstein relation is
satisfied in all heated granular systems where correlations among
particles are weak. On the other side, when excluded volume and energy
dissipation occurring in collisions are increased, non-trivial
correlations appear among positions and velocities of particles. The
global invariant measure cannot be factorized anymore and the response
function depends on it, i.e. on the specific parameters of the
model. As a consequence, the Einstein relation is no more satisfied
and the response function is not trivially predictable. In all
simulations, the decay of $R(t)$ is always faster than that of
$C_1(t)$: this is equivalent to state for the mobility that
\begin{equation}
\mu=\frac{1}{m}\int_0^\infty dt R(t) < \frac{D}{T_g}.
\end{equation}

\begin{center}
\begin{figure}
\includegraphics[width=12cm,clip=true]{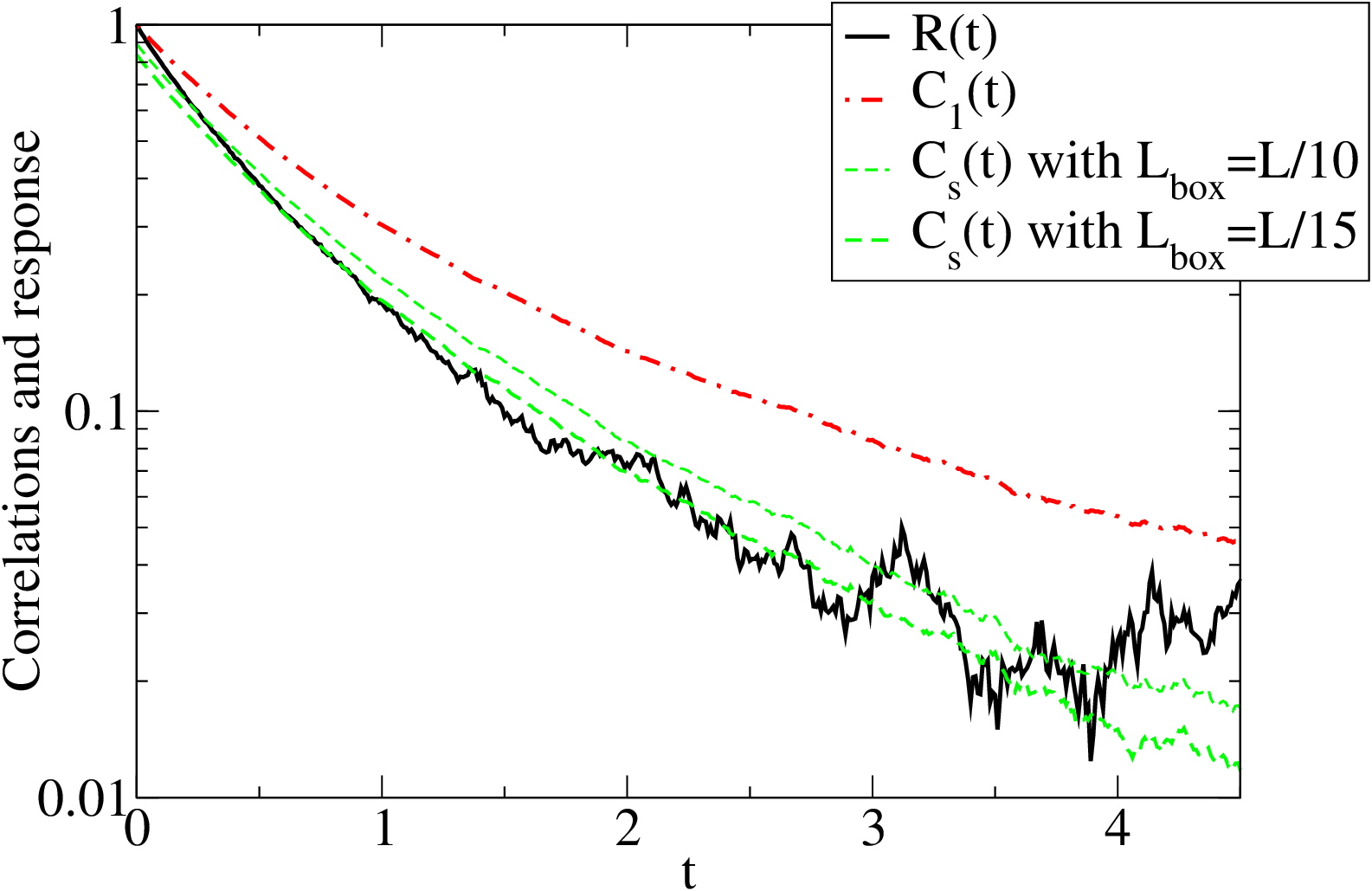}
\caption{Response $R(t)$ and different correlation functions for the
same MD simulation discussed in Figure~\ref{fig:denso}. The normalized
velocity self-correlations $C_1(t)$, as well as the correlation
$C_s(t)$ defined in Eq.~\eqref{eq:risplocal}, for different values of
the coarse graining radius $L_{box}$ are reported.\label{fig:seif}}
\end{figure}
\end{center}

Inspired by a recent work which reported violations of the Einstein
relation in a non equilibrium model~\cite{ss06}, we now conjecture an
effective spatial dependence of the pdf of the velocity component for
a particle at position $\bx$, at time $t$ of the form
\begin{equation} \label{eq:pdflocal}
p_v(v,\bx,t) \sim \exp\left\{-\frac{[v-u(\bx,t)]^2}{2T_g}\right\},
\end{equation}
with $u(\bx,t)$ a local velocity average, defined on a small cell of
diameter $L_{box}$ centered in the particle. Such a hypothesis is
motivated by the fact that, at high density or inelasticities,
spatially structured velocity fluctuations appear in the system for
some time, even in the presence of external
noise~\cite{NETP99,TPNE01}. Following relation~\eqref{eq:fdt} we
propose a formula for the response function to a velocity
perturbation:
\begin{equation} \label{eq:risplocal}
R(t)=C_s=\frac{1}{T_g}\langle v(t) \{v(0)-u[x(0)]\}\rangle.
\end{equation}
Figure~\ref{fig:seif} shows that for small values of the coarse
graining diameter $L_{box}$ (but still large enough to include $5 \div 10$
particles) relation~\eqref{eq:risplocal} is fairly verified. Note
however that the proposed form~\eqref{eq:pdflocal} cannot be exact, a
spatial dependence of $T_g$ should also be included. Furthermore, it
is clear that $\langle u(\bx) \rangle=0$ for any point $\bx$, i.e. the
local velocity field $u(\bx)$ fluctuates in time. Thus, the above
conjecture implies that the characteristic time of variation of these
fluctuations is larger than the characteristic time of response of a
particle: the particle feels, during its response dynamics,
the ``local equilibrium'' average $u(\bx)$. Further investigations
are of course necessary to refine this promising argument.

%%%%%%%%%%%%%%%%%%%%%
\section{Appendix A: the Inelastic Maxwell Model}

The dynamics of a particle in the Inelastic Maxwell Model is
described by the following stochastic process:

\begin{equation} \label{eq:dynamics}
\bv(t+\Delta t)-\bv(t)=\left\{\begin{array}{cc}
\lambda \bv(t) \Delta t  &  \textrm{(with \;\; prob.} \;\; 1-\Delta t)\\
\lambda \bv(t) \Delta t+\Delta \bv_{col}(\bv,\bu,\bsigma)  &  \textrm{(with \;\; prob.} \;\; \Delta t)
\end{array}\right .
\end{equation}
where $\Delta \bv_{col}(\bv,\bu,\bsigma)$ is the effect of an
inelastic collision and has been defined in Eq.~\eqref{eq:colrule},
and $\bu$ is the velocity of the collision partner.

Let us define the two-times covariance matrix $A_{\mu
\nu}(t_1,t_2)=\langle v_\mu(t_1)v_\nu(t_2)\rangle$, with $\mu,\nu \in
\{x,y\}$. Using the evolution law of the system, Eq.~\ref{eq:dynamics}, one can calculate

\begin{eqnarray}
\lefteqn{\frac{\partial A_{\mu\nu}}{\partial t_2}=\lim_{\Delta t_2 \to 0}\left\langle
v_\mu(t_1)\frac{v_\nu(t_2+\Delta t_2)-v_\nu(t_2)}{\Delta t_2}\right\rangle=}\\&=& \langle
v_\mu(t_1)[\lambda v_\nu(t_2)+\Delta v_{col,\nu}]\rangle=\\=\lambda
\langle v_\mu(t_1)v_\nu(t_2)\rangle&-&\frac{1+r}{2}\langle
v_\mu(t_1)\sn[[\bv(t_2)-\bu(t_2)] \cdot \bsigma]\rangle=\\&=&\lambda\langle
v_\mu(t_1)v_\nu(t_2)\rangle\\\lefteqn{-\frac{1+r}{2}\langle v_\mu(t_1)[v_x(t_2)
\sx-u_x(t_2)\sx+v_y(t_2)\sy-u_y(t_2)\sy]\sn\rangle=}\\=\lambda\langle
v_\mu(t_1)v_\nu(t_2)\rangle&-&\frac{1+r}{2}[\langle
v_\mu(t_1)v_x(t_2)\sx\sn\rangle+\langle
v_\mu(t_1)v_y(t_2)\sy\sn\rangle-\\-\langle v_\mu(t_1)u_x\sx\sn\rangle
&-&\langle v_\mu(t_1)u_y\sy\sn\rangle]
\end{eqnarray}
which, assuming absence of correlation between pre-collisional
velocities of different particles and between them and the impact
vector $\bsigma$, gives:
\begin{eqnarray}
\frac{\partial A_{\mu\nu}}{\partial t_2}=\\\lambda\langle v_\mu(t_1)v_\nu(t_2)\rangle-\frac{1+r}{2}[\langle v_\mu(t_1)v_x(t_2)\rangle\langle\sx\sn\rangle+\langle v_\mu(t_1)v_y(t_2)\rangle\langle \sy\sn\rangle]=\\=\kappa A_{\mu\nu}
\end{eqnarray}
where $\kappa=\left(\lambda-\frac{1+r}{4}\right)=-\frac{r(r+1)}{4}$,
and we have used the fact that $\langle
\sm\sn\rangle=\frac{1}{2}\delta_{\mu\nu}$. Since the system is time
translational invariant, we have that
$A_{\mu\mu}(t)=A_{\mu\nu}(0)\exp(\kappa (t_2-t_1))$.

Now, one can perturb at a certain time the velocity $\bv$ of a unique particle, with such
a small perturbation that does not modify the rest of the
system. Starting from~\eqref{eq:dynamics}, one easily computes the
average response to such perturbation:

\begin{equation}
\frac{d \langle \bv \rangle}{dt}=\lambda \langle \bv \rangle - \frac{1+r}{2}\langle \mathbf{F}(\bv,\bu,\bsig)
\rangle=\left(\lambda-\frac{1+r}{4}\right)\langle \bv \rangle,
\end{equation}
where
$\mathbf{F}(\bv,\bu,\bsig)=(v_x\sx^2-u_x\sx^2+v_y\sx\sy-u_y\sy\sx,v_x\sx\sy-u_x\sx\sy+v_y\sy^2-u_y\sy^2)$
and we used the fact that $\langle \bu \rangle=0$. The result is that
the average response decays as the velocity-velocity correlation.

A more general result can be also obtained, starting from the
stochastic evolution equation~\eqref{eq:dynamics}, which can be rephrased as
\begin{equation} \label{eq:linear}
\mathbf{v}(t+\Delta t)=\hat{K}(t) \mathbf{v}(t)+\hat{J}(t)
\end{equation}
where $\hat{K}(t)$ and $\hat{J}(t)$ are two uncorrelated stochastic
(two-dimensional) matrices, with $\langle K_{\mu \nu}(t) \rangle \neq
0$ and $\langle J_{ij}(t) \rangle=0$ . Starting at time $0$ with
$\bv(0)=\bv_0$ and iterating Eq.~\eqref{eq:linear}, one finds that
$\langle \bv(t)|\bv_0 \rangle=\hat{L}^t \bv_0$ where
$L_{\mu\nu}=\langle K_{\mu\nu}\rangle$. From
relation~\eqref{eq:angelo} (generalized to two dimensions), the
proportionality of all the correlation functions follows.

\vspace{.5cm}

%%%%%%%%%%%%%%%%%%%%%%%%
\noindent {\it Acknowledgments.--} We wish to thank F. Cecconi,
M. Cencini, L. Leuzzi and U. Marini Bettolo Marconi for discussions
and a critical reading of the manuscript.

%%%%%%%%%%%%%%%%%%%%%%%%
%%%%%%%%%%%%%%%%%%%%%%%%
%%%%%%%%%%%%%%%%%%%%%%%%
\section*{References}

\bibliographystyle{unsrt}
\bibliography{fluct.bib}

\end{document}